\documentclass[11pt,a4paper]{article}
\usepackage{graphicx}
\usepackage{subfigure}
\usepackage{epsfig}
\newcommand{\beq}{\begin{equation}}
\newcommand{\eeq}{\end{equation}}
\newcommand{\ba}{\begin{array}}
\newcommand{\ea}{\end{array}}
\newcommand{\beqa}{\begin{eqnarray}}
\newcommand{\eeqa}{\end{eqnarray}}
\newcommand{\bd}[1]{ \mbox{\boldmath $#1$}  }
\newcommand{\oh}{{1\over 2}}

\newcommand{\nn}{\nonumber}
\newcommand{\APNY}[1]{Ann. Phys. (N.Y.) {\bf {#1}}}

\newcommand{\NPA}[1]{Nucl. Phys. {\bf A{#1}}}

\begin{document}
\input{epsf.sty}

\title{CLUSTER EXPANSION OF COLD ALPHA-MATTER ENERGY}

\author{F. Carstoiu$^1$, \c S. Mi\c sicu$^1$, V. B\u al\u anic\u a$^1$ and  
M. Lassaut$^2$,\\
$^1$ National Institute for Nuclear Physics and Engineering,\\ 
P.O.Box MG-6, RO-077125 Bucharest-Magurele, Romania\\
\vspace{1mm}\\
$^2$  Institut de Physique Nucl\'eaire\\
  IN2P3-CNRS, Universit\'e Paris-Sud 11\\
  F-91406 Orsay Cedex, France}

\date{}
\maketitle
\begin{center}
(Received \today )
\end{center}

\begin{abstract}
In the cluster expansion framework of Bose liquids we calculate analytical expressions
of the two-body, three-body and four-body diagrams
contributing to the g.s. energy of an infinite system of neutral alpha-particles at 
zero-temperature, interacting via the strong nuclear forces exclusively. This is  analytically tractable by
assuming  a density dependent two-body correlation function  of Gaussian type. For the $\alpha-\alpha$  
potential we adopt the phenomenological Ali-Bodmer interaction  and semi-microscopic potentials 
obtained from the Gogny force parametrizations.
We show that under such assumptions we achieve a rapid convergence in the cluster
expansion, the four-body contributions to the energy being smaller than the 
two-body and three-body contributions by at least an order of magnitude.
\end{abstract}
\bigskip

{\em Key words:\/} Equation of state, nuclear matter, supernova explosion, cluster 
expansion, quantum liquids.

{\section{INTRODUCTION \label{int} }}

 Understanding the properties of $\alpha$ matter has retained a lot of 
attention in recent years. This situation is mainly due to the believe 
that this type of hadronic matter 
occurs in astrophysical environment in deconfined form. 
In the debris of a supernova explosion, a substantial fraction of hot and dense matter 
resides in $\alpha$ particles and therefore the equation of state of  matter at subnuclear densities
is essential in simulating the supernova collapse and explosions and is also 
important for the formation of the supernova neutrino signal \cite{lattimer}.

The aim of the present work is to investigate the equation of state of $\alpha$ matter 
from the standpoint of the cluster expansion method of Bose liquids. 
We consider a cold ($T$=0) system of $\alpha$-particles interacting only by means 
of the strong nuclear force. Similarly to the case of ordinary nuclear matter (composed of 
protons and neutrons) the Coulomb interaction is switched-off.
The internal structure of the $\alpha$ clusters is accounted only in the 
determination of the $\alpha-\alpha$ potential, the single particle structure being 
incorporated in the cluster densities that are folded with the effective nucleon-nucleon
 ($NN$) interaction.   
 In continuation to the previous assumption, no Pauli blocking effects are included. 
Naturally, since the constituents of the $\alpha$ particles are fermions, one should expect 
that the Pauli principle is manifest when two or more $\alpha$ clusters start to overlap.  
As revealed by the work of R\"opke and collab. \cite{roepke} one should expect from the 
action of this principle a dissolution of the $\alpha$ cluster in protons and
neutrons above the so-called Mott density, which presumably lays between a fifth and a third of the
nuclear matter saturation density.
It is well known \cite{schafer}, that two-body correlation functions (TBCFN) obtained 
by minimization of the energy functional truncated 
at the lowest order may lead to an unphysical deep minimum.   
 We adopt the cluster expansion method of a Bose liquid \cite{aviles58} and
use the simple Jastrow ansatz
involving state-independent two-body correlation functions . These are taken in a Gaussian form, without
overshooting near the healing distance,
\beq
 f(r)=1-e^{-\beta^2r^2} \ .
 \label{eq1} 
 \eeq
The parameter $\beta$ is determined from the normalization condition for the correlation 
function \cite{fe69}
\beq
 4\pi\rho \int_0^\infty dr r^2(f^2(r)-1)=-1 \ .
\label{eq2}
\eeq
This condition ensures that the mean square deviation of the correlation
function from unity is a small quantity and has an exponential healing. Consequently the 
dependence  $\beta=\beta(\rho)$ , where 
$\rho={N_\alpha}/{V}$ is the density of $\alpha$-particles, reads,
\beq 
\beta 
=\sqrt{\pi}\left [ \rho\left ( 2-\frac{1}{2\sqrt{2}}\right )\right ]^{1/3} \ .
\label{eq3}
\eeq   
 For the $\alpha-\alpha$ interaction, we adopt two types of Gaussian-like
potentials containing a short range soft repulsive part and a long range shallow attractive part.
 The first one 
is the $S$-state Ali-Bodmer (AB) potential \cite{AB66}:
\beq
\label{eq4}
 v(r)=V_Re^{{-\mu_R^2}r^2}-V_Ae^{{-\mu_A^2}r^2} \ .
\eeq
 In this expression, $V_A=130$ MeV, $V_R=475$ MeV, $\mu_A=$0.475 fm$^{-1}$, $\mu_R=$0.7 fm$^{-1}$. 
This potential obtained by a  fit  of the low energy $\alpha-\alpha$ phase shifts, can be 
considered as an approximation to the supersymmetric partner of the deep potential of Buck et al. \cite{buck}.
The second type of potential is a sum of three Gaussian and is derived from two recent parametrizations
of the Gogny effective $N-N$ force \cite{meyer}. Two explicit forms, labeled (D1) and (D1N),
are given in \cite{misicarst}. We have checked that these potentials satisfying the 
integrability condition $\int_0^{\infty} r \vert v(r) \vert dr < \infty$ are in agreement with 
the Levinson theorem 
in the sense given in 
\cite{neudacin}. In the absence of the Coulomb interaction, the $\alpha-\alpha$ interactions (AB) and (D1N)
provide  a weakly bound g.s. $J^{\pi}=0^+$.  In contrast to the aforementioned  interactions, the   $\alpha-\alpha$ (D1) 
interaction is characterized by the absence of bound states.

In section {\bf 2} we  apply the cluster expansion method to our problem and express 
all terms of the development in a compact form. Our results are discussed in section  {\bf 3}.

\vspace{1cm}

\section{CLUSTER EXPANSION METHOD}

\begin{figure}[h]
\centerline{\includegraphics[width=10cm]{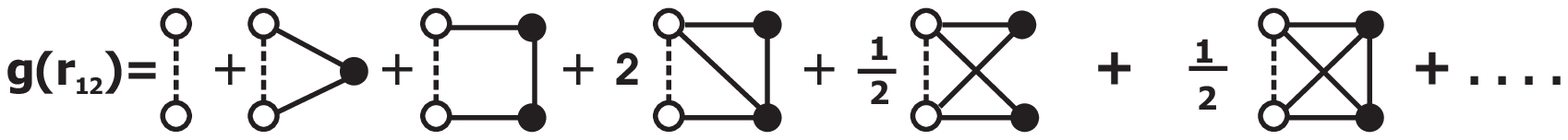}}
\caption{Diagrammatic expansion of the radial distribution function. There are 4
independent diagrams contributing to the 4th order term: ring, diagonal, opened and
connected.}
\label{fig1}
\end{figure}

Let the energy of a system of strongly interacting $N$-bosons
in the Jackson-Feenberg form \cite{Cla66},
\beq
E=\oh \rho N\int d\bd{r} g(r) v^*(r) \ ,
\label{eq5} 
\eeq 
where $g(r)$ is the radial distribution function and $v^*$ is the
 effective Jackson-Feenberg potential, 
\beq
\label{eq6}
 v^*(r)=v(r)-\frac{\hbar^2}{2m_{\alpha}}\nabla^2 \ln f(r) \ .
\eeq
In the case of $\alpha$-matter the energy Eq. (\ref{eq5}) is measured relative to
the rest energy of a free $\alpha$-particle. Using Eqs. (\ref{eq1}) and (\ref{eq4}) we obtain
\beq
  v^*(r)={\frac{2 c \beta ^2 \left(3-e^{r^2 \beta ^2} \left(3-2 r^2 \beta ^2\right)\right)}{\left(1-e^{r^2 \beta
^2}\right)^2}+V_Re^{-r^2 \mu
_R^2} -V_Ae^{-r^2 \mu _A^2} } \ ,
\label{eq7}
\eeq
where $c={\hbar^2}/{2m_\alpha}$. The main trick allowing the   cluster expansion of the above expression consists
in expanding the radial distribution function in powers of the  small parameter $\omega=\int d\bd{r} h(\bd{r})$.
It uses   the fact that the function $h=f^2-1$ is of short-range.
Accordingly, the cluster expansion of $g(r)$ reads,
\newpage
\beqa
\label{eq8}
g(\bd{r}_{12})&=&f^2(\bd{r}_{12})\left\{ 1+
\rho \int d\bd{r}_3 h(r_{13})h(r_{23})\right. \nn\\
&+& \rho^2\int d\bd{r}_3\int d\bd{r}_4 \bigg[ 
h(r_{13})h(r_{24})h(r_{34})+2h(r_{13})h(r_{24})h(r_{34})h(r_{14})\nn\\
&+& \frac{1}{2} h(r_{13})h(r_{23})h(r_{14})h(r_{24})\nn\\&+& \left.
\frac{1}{2}h(r_{13})h(r_{14})h(r_{23})h(r_{24})h(r_{34}) \bigg]
+{\cal O}(\rho^3) \right\} \ .
\eeqa
A diagrammatic expansion of the radial distribution function is depicted in Fig.
\ref{fig1}. For a given $n$-body diagram there are field points (open circles) and
dummy points (filled circles). For each dummy point there is an integration
$\rho\int d\vec r_i$. A bond between open points involves a factor $
f^2$ in the integrand. Any other bond implies a factor $h=f^2-1$ in the
integrand.
Mutatis mutandis, the ground state energy per $\alpha$-particle is then 
expanded in powers of the density :
\beq E=E_2+E_3+E_4+\ldots
\label{eq9}  
\eeq
Above $E_n$ stands for the contribution to the energy per particle arising 
from $n$-body diagrams. In what follows we calculate these first three terms
of the expansion.

Partial contributions to the total energy are displayed using the  $\alpha-\alpha$ Ali-Bodmer interaction (Eq.(\ref{eq4})).
  
{\subsection{Two-body diagram $E_2$ \label{e2} }}

Let us split $E_2$ only into kinetic and potential components :
\beq E_{\rm 2K}=\frac{1}{2}\frac{\rho}{\Omega}\int d\bd{r}_1d\bd{r}_2 f^2(r_{12})
\left[-\frac{\hbar^2}{2m_\alpha}\nabla^2\ln f(r_{12})\right] \ ,
\label{rq10} 
\eeq
\label{eq11}
\beq E_{\rm 2V}=\oh \frac{\rho}{\Omega}\int d\bd{r}_1d\bd{r}_2f^2(r_{12})v(r_{12}) \ ,
\eeq
where $\Omega$ is the integration volume. Consider the general 6-dimensional integral
\beq I_2=\oh \frac{\rho}{\Omega}\int d\bd{r}_1d\bd{r}_2 p(r_{12}) \ ,
\label{eq12}
\eeq
where $\bd{r}_{12}=\bd{r}_1-\bd{r}_2$ and $p$ a generic function. 
With the unitary transformation (unit Jacobian), 
\beq \bd{r}_1=\bd{R}+\oh\bd{s} \ , \nn \eeq
\beq \bd{r}_2=\bd{R}-\oh\bd{s} \ ,\nn \eeq
the  integral $I_2$  (Eq.(\ref{eq12})) reads:
\beq I_2=\oh\frac{\rho}{\Omega}\int d\bd{R}d\bd{s}~p(s)=2\pi \rho \int_0^\infty 
d{s}~s^2 p(s) \ .\eeq
Note that integration over the c.m. variable $\bd R$ gives the integration
volume $\Omega$. Therefore  we have
\beqa 
E_{\rm 2K}&=&-\oh \rho \int d\bd{s}~f^2(s)\left[\frac{\hbar^2}{2m_\alpha}\nabla^2 \ln f(s)\right]\nn\\
&=&-\oh \rho \int d\bd{s}~f^2(s)\frac{\hbar^2}{2m_\alpha}\left[f(s)\nabla^2 f(s)-{(\nabla f(s))}^2\right] \ ,\eeqa
\label{eq13}
\beq E_{\rm 2V}=\oh \rho\int d\bd{s}~ f^2(s)v(s)  \ .
\eeq
\begin{figure}[h]
\centerline{\includegraphics[width=10cm]{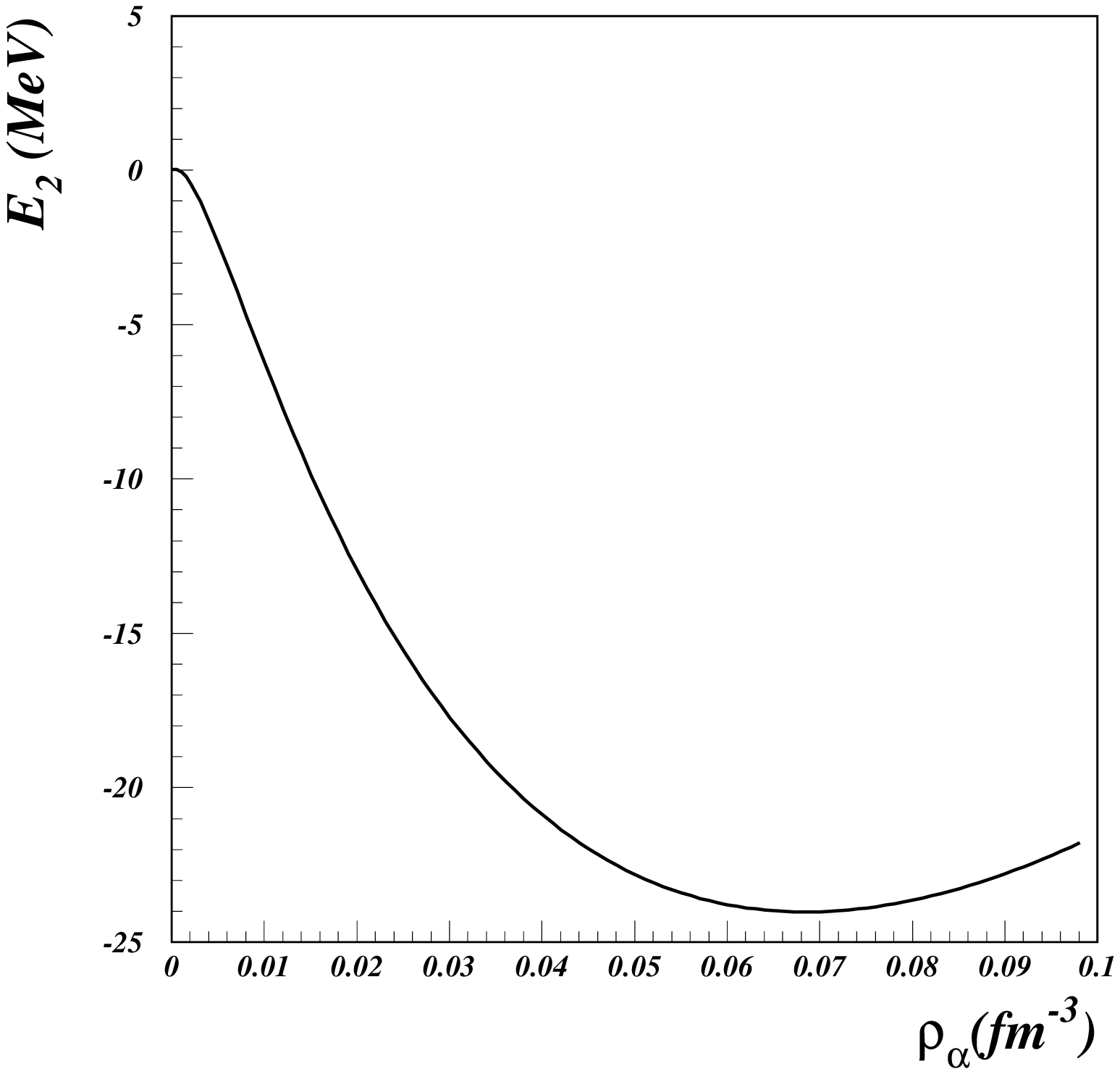}}
\caption{Density dependence of the ${\it E}_2$ component.
\label{fig2}}
\end{figure}
Summing up these two contributions we obtain,
\beq  E_2=\frac{1}{2} \rho \int d\bd{r} f^2(r) v^*(r)  \ ,
\label{eq14}
\eeq
or in analytical form
\beq E_2={\frac{1}{4} \pi ^{3/2} \rho  }
{\left(\frac{3 \sqrt{2} c}{\beta }+2V_RF_R-2V_AF_A\right)} \ .
\label{eq15} 
\eeq
Defining the auxiliary function
\beq  \chi_{ij}(\beta,\mu)=\frac{1}{(i\beta^2+j\mu^2)^{\frac{3}{2}}} \ ,
\label{eq16}
\eeq
we then have 
\beq F_{i}=-\chi_{0,1}(\beta,\mu_{i})+2\chi_{1,1}(\beta,\mu_{i})
-\chi_{2,1}(\beta,\mu_{i}) \ ,\eeq
where $(i=A,R)$. The dependence of $E_2$ on density is displayed in Fig. \ref{fig2}. 
We observe that already the component $E_2$
has a shallow minimum at a density almost two times 
the saturation density of normal nuclear matter. This is in contrast to the result of
ref. \cite{misicarst}, where it has been shown that TBCFN's obtained from Pandharipande-Bethe equation
lead to a collapse of $E_2$ component. This effect arises entirely from the
density dependence of our particular functional form of TBCFN and its
derivatives.

\vspace{1cm}

{\subsection{Three-body diagram $E_3$}\label{E3}}

\begin{figure}
\centering
\mbox{\subfigure{\epsfig{figure=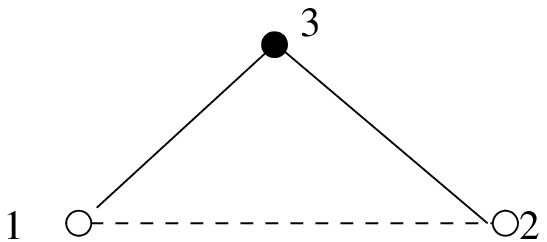,width=0.50\textwidth}}
      \subfigure{\epsfig{figure=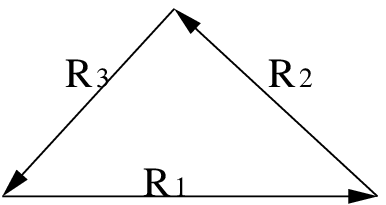,width=0.38\textwidth}}
      }
\caption{${\it E}_3$ diagram and $\delta$-diagram
}
\label{fig3}
\end{figure}

The diagram corresponding to the three-body energy is given
on the left panel of Fig. \ref{fig3}. By definition,
\beq E_{\rm 3K}=\oh \left(\frac{\rho^2}{\Omega}\right)\int d\bd{r}_1d\bd{r}_2d\bd{r}_3f^2(r_{12})h(r_{23})h(r_{31}) \left[-\frac{\hbar^2}{2m_\alpha}\nabla_1^2 \ln f(r_{12})\right] \ ,\eeq
\beq 
E_{3V}=\oh \left(\frac{\rho^2}{\Omega}\right)\int d\bd{r}_1d\bd{r}_2d\bd{r}_3f^2(r_{12})h(r_{23})h(r_{31}) v(r_{12})  \ .
\eeq
Like previously for the $E_2$    component we consider the following generic integral
\beqa I_3&=&\oh \left(\frac{\rho^2}{\Omega}\right)\int d\bd{r}_1d\bd{r}_2d\bd{r}_3p_1(r_{12})p_2(r_{23})p_3(r_{31})\nn\\
&=&\oh \left(\frac{\rho^2}{\Omega}\right)\int \prod_1^3 d\bd{r}_i \prod_1^3 d\bd{R}_i p_1(R_1)p_2(R_2)p_3(R_3)\nn\\
&~&\times\delta(\bd{R}_1-\bd{r}_1+\bd{r}_2)
\delta(\bd{R}_2-\bd{r}_2+\bd{r}_3)\delta(\bd{R}_3-\bd{r}_3+\bd{r}_1) \ . \eeqa
Integration of $\bd r_i$ variables is straightforward and we obtain,
\beq I_3=\oh \rho^2\int d\bd{R}_1d\bd{R}_2d\bd{R}_3p_1(R_1)p_2(R_2)p_3(R_3) \delta(\bd{R}_1+\bd{R}_2+\bd{R}_3) \ .
\label{eq25}
\eeq
Eq.(\ref{eq25}) is useful for both numerical and analytical integration since
the angular dependence is isolated in a $\delta$ function as shown in the
right panel of Fig. \ref{fig3}. Introducing the 
double-folding $\delta$-integral,
\beq V_{p,q}^{\delta}(\bd{R}_1)=\int d\bd{R}_2d\bd{R}_3p(R_2)q(R_3)\delta(\bd{R}_1+\bd{R}_2+\bd{R}_3) \ ,\label{eq23}\eeq
the integral $I_3$ is reduced to, 
\beq I_3=\oh \rho^2\int d\bd{R}_1p_1(R_1)V_{p2,p3}^{\delta}(\bd{R}_1) \ .\eeq
Applying this technique to the $E_3$ term, we have,
\beq E_{\rm 3K}=\oh \rho^2\int d\bd{R}_1
f^2(R_1)\left(-\frac{\hbar^2}{2m_\alpha}\right)\left[f(R_1)\nabla^2f(R_1)-{\left(\nabla
f(R_1)\right)}^2\right]V_{h,h}^{\delta}(R_1) \label{eqa1} \eeq
\beq E_{\rm 3V}=\oh \rho^2\int d\bd{R}_1f^2(R_1)v(R_1)V_{h,h}^{\delta}(R_1) \ .\eeq
Consequently the total three-body energy contribution reads,
\beq E_3=E_{\rm 3K}+E_{\rm 3V}=\oh \rho^2 \int  d\bd{R}_1v^*(R_1)f^2(R_1)V_{h,h}^{\delta}(R_1) \ , \label{eqa2}\eeq
where we have defined according to the prescription (\ref{eq23})\\
\beq V_{h,h}^{\delta}(\bd{R}_1)=\int d\bd{R}_2d\bd{R}_3h(R_2)h(R_3)\delta(\bd{R}_1+\bd{R}_2+\bd{R}_3) \ .\label{eq23p}\eeq
For our particular selection of the TBCFN,
\beq V_{h,h}^\delta(R)=\frac{\pi^{3/2}}{\beta^3} 
\left(\frac{1}{8} e^{-R^2 \beta ^2} - \frac{4 \sqrt{3}}{9} e^{\frac{-2 R^2 \beta ^2}{3}}+ \sqrt{2} e^{\frac{-R^2 \beta ^2}{2}}\right)  \ ,\label{eq49} \eeq
and
\beq{E_3=\frac{\pi ^3 \rho ^2}{\beta ^3}}
 \left[ \frac{c}{\beta} \left(-\frac{27}{32 \sqrt{2}}-\frac{29}{24 \sqrt{3}}+ \frac{84}{25 \sqrt{5}} \right) +
 V_RG_R-V_AG_A \right] \ ,\eeq
where for $(i=A,R)$,
\beqa G_i&=&\frac{1}{16} [\chi_{1,1}(\beta,\mu_i)-2\chi_{2,1}(\beta,\mu_i)+\chi_{3,1}(\beta,\mu_i)  
+ 32(\chi_{1,2}(\beta,\mu_i)-2\chi_{3,2}(\beta,\mu_i)\nn\\&+&\chi_{5,2}(\beta,\mu_i)-\chi_{2,3}(\beta,\mu_i)+2\chi_{5,3}(\beta,\mu_i)-
\chi_{8,3}(\beta,\mu_i)) ].\nn\\
\eeqa
\begin{figure}[h]
\centerline{\includegraphics[width=10cm]{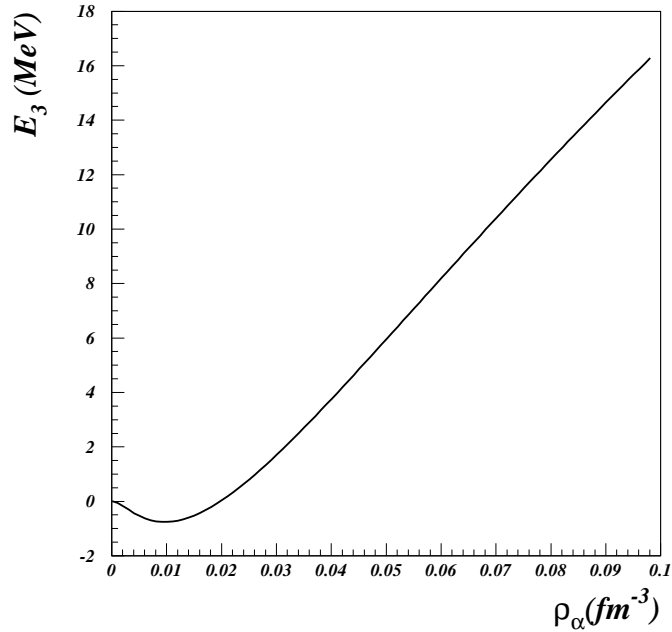}}
\caption{Density dependence of the ${\it E}_3$ component.
\label{fig4}}
\end{figure}

The density dependence of the three-body energy is given in Fig. \ref{fig4}.
We notice that contrary to $E_2$ which is attractive, $E_3$ is only weakly
attractive at low density and becomes strongly repulsive with increasing density.  

\vspace{1cm}

\begin{figure}
\centering
\mbox{\subfigure{\epsfig{figure=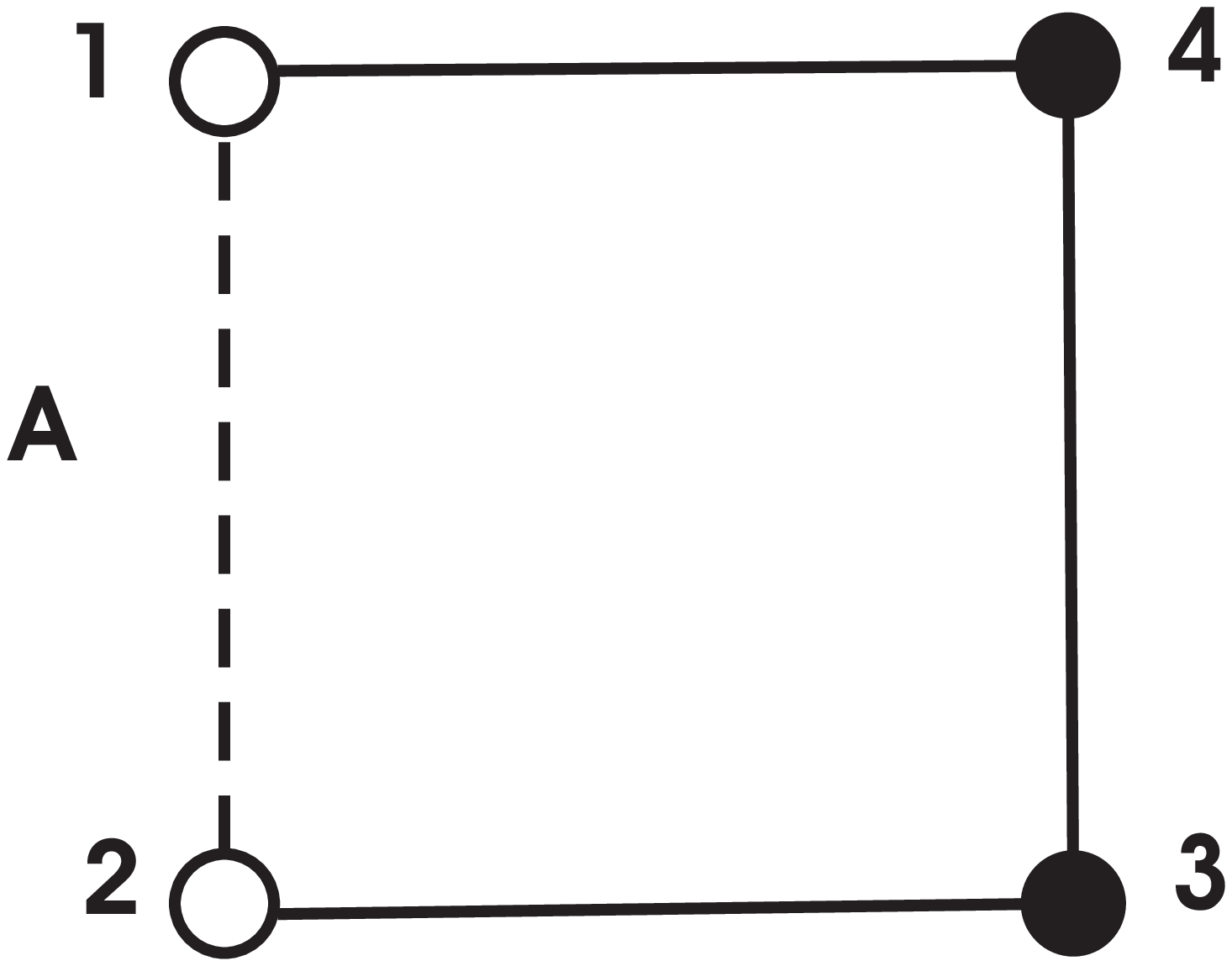,width=0.30\textwidth}}
      \subfigure{\epsfig{figure=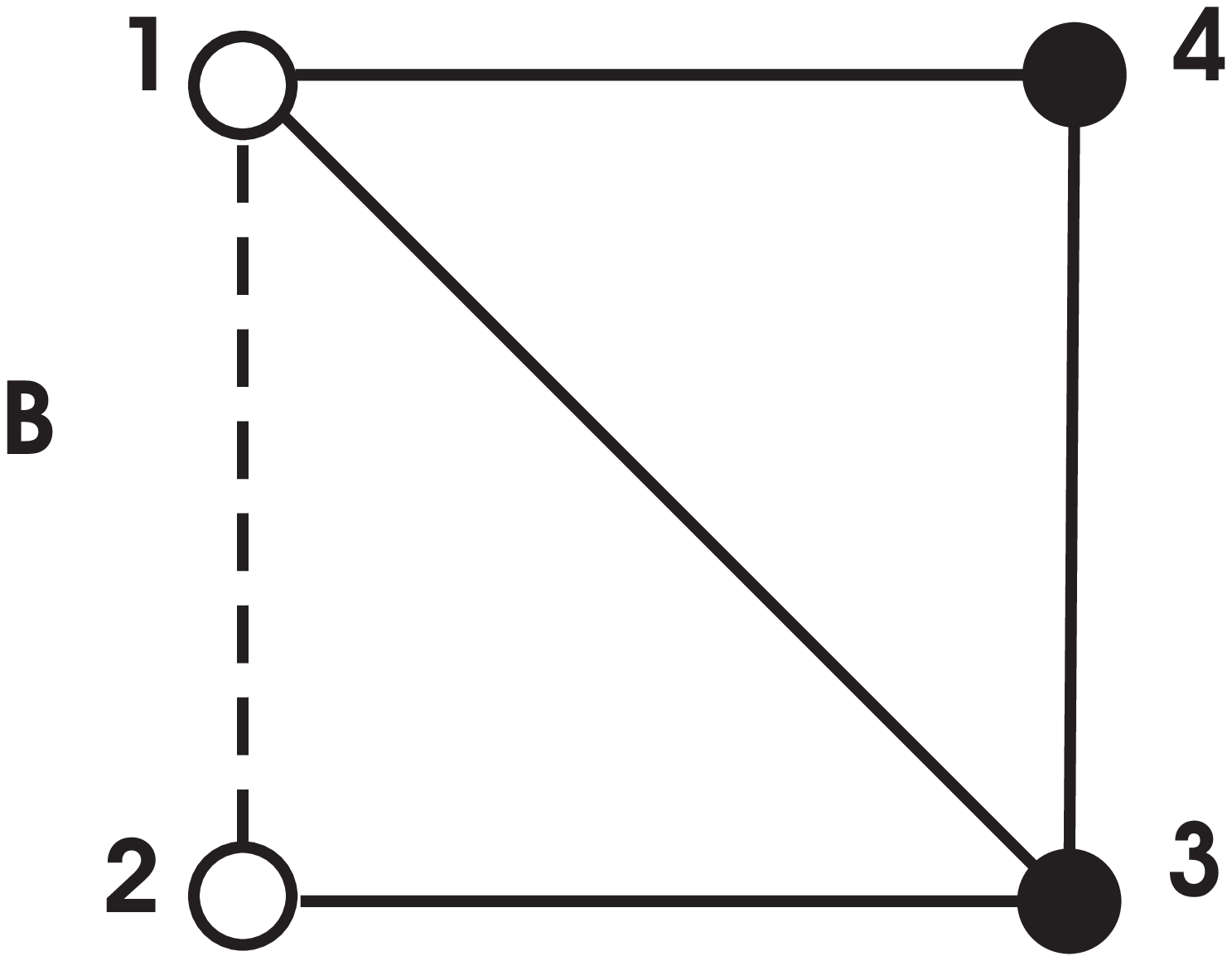,width=0.30\textwidth}}
     }
\mbox{\subfigure{\epsfig{figure=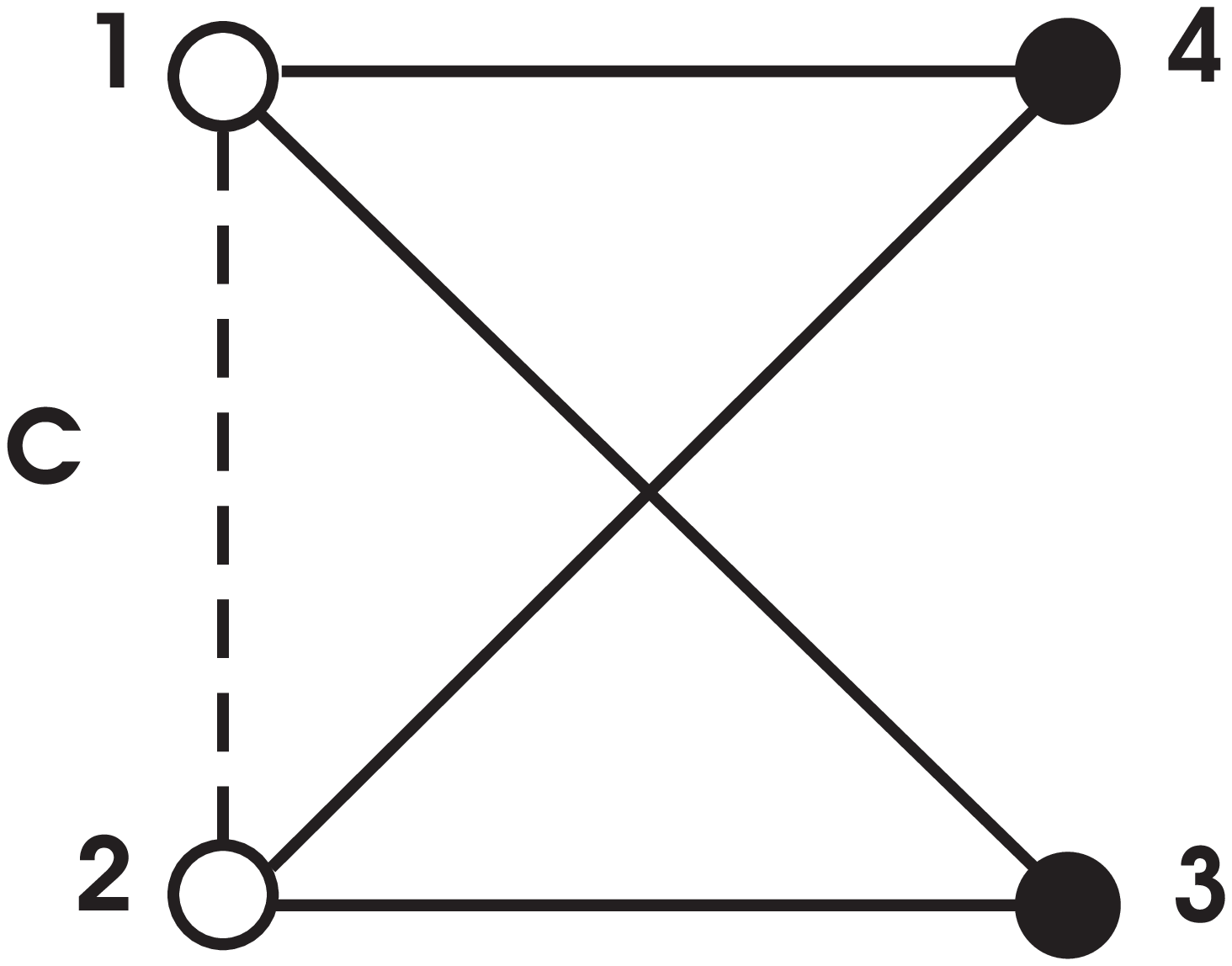,width=0.30\textwidth}}
      \subfigure{\epsfig{figure=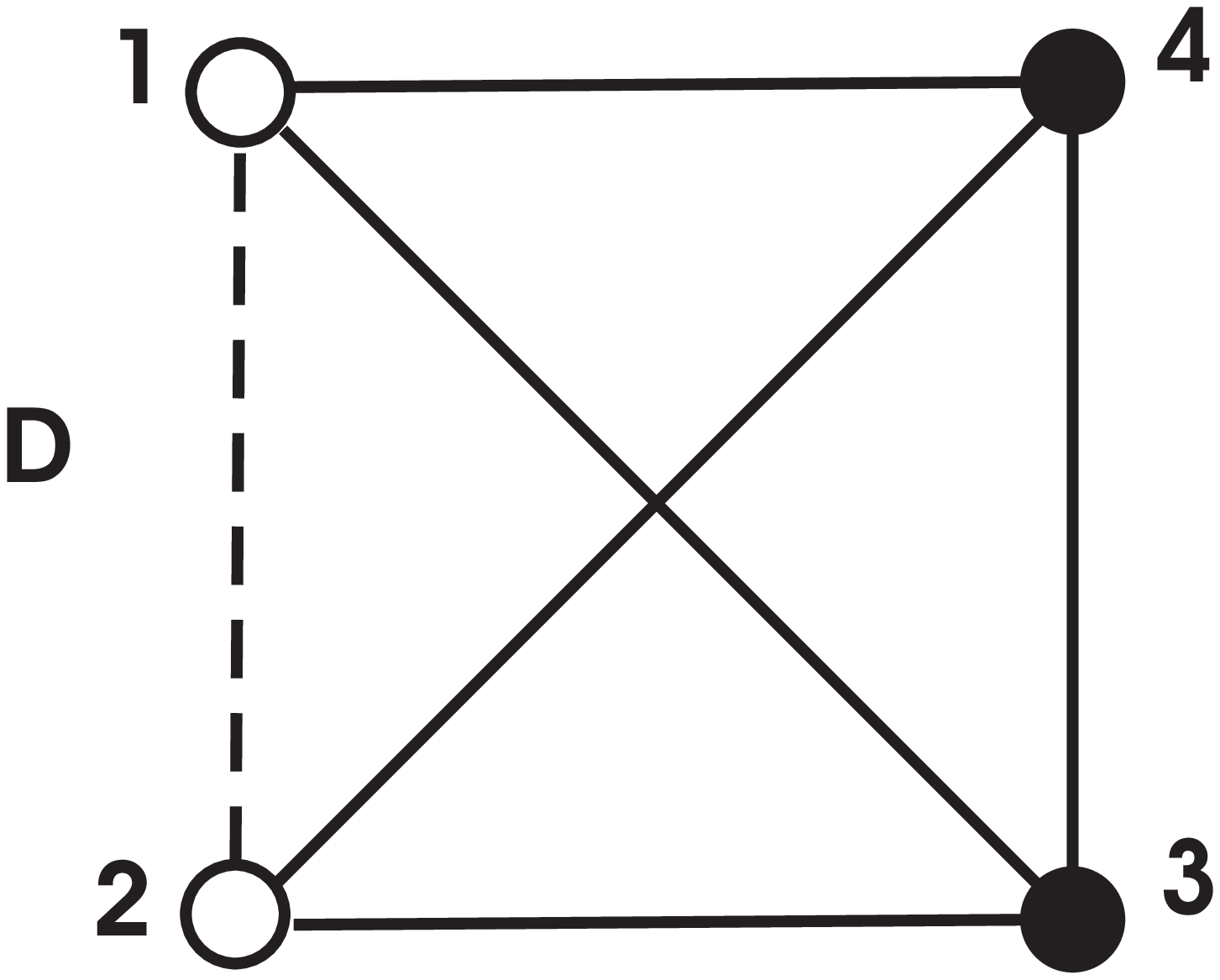,width=0.30\textwidth}}
      }
\caption{(A) Ring, (B) diagonal, (C) opened and (D) connected diagrams contributing to the ${\it E}_4$ component. 
}
\label{fig5}
\end{figure}

{\subsection{The four-body diagram $E_4$}\label{e4}}
 The diagrams contributing to the $E_4$ component are
depicted  in Fig.
 \ref{fig5}. They are dubbed as ring, diagonal, opened and connected diagrams.
Explicit expressions of these components are, 
\beq E_{\rm 4R}=\oh \frac{\rho^3}{\Omega}\int \prod_1^4d\bd{r}_iv^*(r_{12})f^2(r_{12})h(r_{23})h(r_{34})h(r_{41
}) \ ,\eeq
\beq E_{\rm 4D}=2\oh \frac{\rho^3}{\Omega}\int \prod_1^4d\bd{r}_iv^*(r_{12})f^2(r_{12})h(r_{23})h(r_{34})h(r_{41
})h(r_{24}) \ ,\eeq
\beq E_{\rm 4O}=\oh \oh \frac{\rho^3}{\Omega}\int \prod_1^4d\bd{r}_iv^*(r_{12})f^2(r_{12})h(r_{23})h(r_{41})h(r_{24
})h(r_{13}) \ ,\eeq
\beq E_{\rm 4C}=\oh \oh \frac{\rho^3}{\Omega}\int \prod_1^4d\bd{r}_iv^*(r_{12})f^2(r_{12})h(r_{23})h(r_{34})h(r_{41
})h(r_{24})h(r_{13}) \ .\eeq
\vspace{1cm}
{\subsubsection{Calculation of $E_{\rm 4R}$ ring diagram} \label{E4r}}
\begin{figure}[h]
\centerline{\includegraphics[width=10cm]{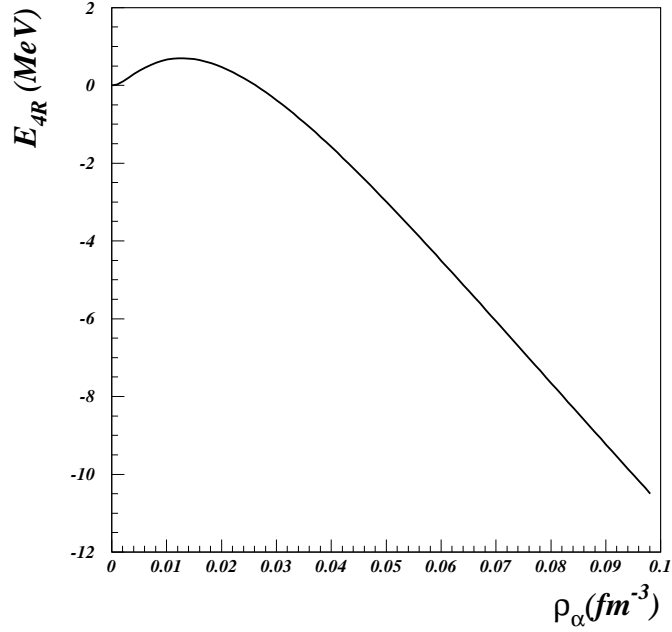}}
\caption{Density dependence of the ${\it E}_{\rm 4R}$ component
for the AB potential.
\label{fig6}}
\end{figure}
It is easy to see that the ring diagram contribution to the energy is given by,
\beq E_{\rm 4R}=\frac{1}{2}\rho^3 \int \prod_1^4 d\bd{R}_if^2(R_1)v^*(R_1)h(R_1)h(R_2)h(R_3)h(R_4)\delta(\bd{R}_1+\bd{R}_2+\bd{R}_3+\bd{R}_4) \eeq
or equivalently,
\beqa E_{\rm 4R}&=&\frac{1}{2}\rho^3 \int \prod_1^2 d\bd{R}_if^2(R_1)v^*(R_1)h(R_2)V_{h,h}^{\delta}(\bd{R}_1+\bd{R}_2)\nn\\
&=&\frac{1}{2}\rho^3 \int \prod_1^3 d\bd{R}_if^2(R_1)v^*(R_1)h(R_2)V_{h,h}^{\delta}(R_3)\delta(\bd{R}_3-\bd{R}_1-\bd{R}_2),\eeqa
in terms of the function $V_{h,h}^{\delta}$ defined in Eq.(\ref{eq23p}).  
In the hierarchy of $\delta$ kernel let us introduce, 
\beq V^{\delta\delta}_{p,q,r}(R_1)=\int d\bd{R}_2 d\bd{R}_3p(R_2)V_{q,r}^{\delta}(R_3)\delta(\bd{R}_3-\bd{R}_1-\bd{R}_2) \ .\eeq
The final integral is simply,
\beq E_{\rm 4R}=\oh \rho^3 \int d\bd{R}_1 v^*(R_1)f^2(R_1)V^{\delta\delta}_{h,h,h}(R_1) \ .\eeq
Using the expression for $V_{h,h}^{\delta}$ given in  Eq.(\ref{eq49}) we obtain the expression of
 $V^{\delta\delta}_{h,h,h}$,
\beqa  V^{\delta\delta}_{h,h,h}(R)&=&\frac{\pi ^3}{\beta^6} 
\left(\frac{1}{24 \sqrt{3}}e^{-\frac{2}{3} R^2 \beta ^2} -\frac{3}{8 \sqrt{2}} e^{-\frac{R^2 \beta ^2}{2}} \right. \nn\\ 
 &+&\left. \frac{12}{5 \sqrt{5}} e^{\frac{-2 R^2 \beta^2}{5}}-\frac{8}{3 \sqrt{3}} e^{\frac{-R^2 \beta ^2}{3}}\right)  \ .\eeqa
The corresponding expression  for $E_{\rm 4R}$ is:
\beqa E_{\rm 4R}=\frac{\pi ^{9/2} \rho ^3}{ \beta ^6}\left [ \frac{c}{\beta}
\left( \frac{3}{4} + \frac{3}{128  \sqrt{2}}+\frac{7}{4 \sqrt{3}}+\right. \right.\nn\\ \left. \left. -\frac{12}{25 \sqrt{5}}-
\frac{240}{49  \sqrt{7}} \right)+H_RV_R-H_AV_A \right] \label{momo1} \ . \eeqa
  In the equation (\ref{momo1}) the $H_i$'s are defined by 
\beqa H_i=\frac{1}{16} \left(-6\chi_{1,2}(\beta,\mu_i)+12\chi_{3,2}(\beta,\mu_i)-6
\chi_{5,2}(\beta,\mu_i)-64\chi_{1,3}(\beta,\mu_i)+\right.\nn\\ \left.\chi_{2,3}(\beta,\mu_i)+128\chi_{4,3}(\beta,\mu_i)-
2\chi_{5,3}(\beta,\mu_i)-64\chi_{7,3}(\beta,\mu_i)\right.\nn\\ \left.
+\chi_{8,3}(\beta,\mu_i)+96\chi_{2,5}(\beta,\mu_i)-192\chi_{7,5}(\beta,\mu_i)+96\chi_{12,5}(\beta,\mu_i)\right) \ ,\nn\eeqa
where $(i=A,R)$.

From the inspection of Fig.\ref{fig6} we infer that the ring contribution to the
four-body energy has the same behavior as $E_3$ save for a factor -1.
\vspace{1cm}

{\subsubsection{Calculation of $E_{\rm 4D}$ diagonal diagram} \label{E4d}}
In order to isolate the angular variables, we proceed as follows,
\beqa E_{\rm 4D}&=&2\oh \frac{\rho^3}{\Omega}\int \prod_1^4 d\bd{r}_iv^*(r_{12})f^2(r_{12})h(r_{23})h(r_{34})h(r_{41})h(r_{24})\nn\\
 E_{\rm 4D} &=&2\oh \frac{\rho^3}{\Omega}\int \prod_1^5 d\bd{R}_i\prod_1^4 d\bd{r}_iv^*(R_1)f^2(R_1)h(R_2)h(R_3)h(R_4)h(R_5) \nn\\
&~&\times\delta(\bd{R}_1-\bd{r}_1+\bd{r}_2)\delta(\bd{R}_2-\bd{r}_2+\bd{r}_3)\delta(\bd{R}_3-\bd{r}_3+\bd{r}_4)\nn\\
&~&\times\delta(\bd{R}_4-\bd{r}_4+\bd{r}_1)\delta(\bd{R}_5-\bd{r}_2+\bd{r}_4) \ .\eeqa
Performing the same manipulations  as above we obtain,
\beqa E_{\rm 4D}=2\cdot\frac{1}{2}\rho^3 \int \prod_1^5 d\bd{R}_if^2(R_1)v^*(R_1)h(R_2)h(R_3)h(R_4)h(R_5)\nn\\\delta(\bd{R}_5+\bd{R}_1+\bd{R}_4) \delta(\bd{R}_5-\bd{R}_2-\bd{R}_3) \ .\eeqa

We observe that variables $\bd R_i, i=1,2,3,4$ are decoupled. It can be seen easily that: \\
\beq E_{\rm 4D}=\rho^3 \int d\bd{R}_1d\bd{R}_4d\bd{R}_5f^2(R_1)v^*(R_1)h(R_4)h(R_5) V^{\delta}_{h,h}(R_5) \delta(\bd{R}_5+\bd{R}_1+\bd{R}_4)  \ .\eeq
Finally we have,
\beqa E_{\rm 4D}=\frac{\pi ^{9/2} \rho ^3}{288 \beta ^6} \left(\frac{c}{\beta}\left(\frac{27   }{8 }-\frac{26624 \sqrt{3}-7344  }{49\sqrt{7} }+\frac{2432
\sqrt{6}-10368}  {25\sqrt{5} }-
\right.\right.\nn\\ \left.\left.{\frac{945  }{2 \sqrt{2} } 
+74 \sqrt{3} }
-{\frac{15552  }{121 \sqrt{11}  }+\frac{580608  }{169 \sqrt{13}  }
-\frac{442368 
}{361 \sqrt{19} }}\right)+V_RD_R-V_AD_A\right) \ ,
\eeqa

\begin{figure}[h]
\centerline{\includegraphics[width=10cm]{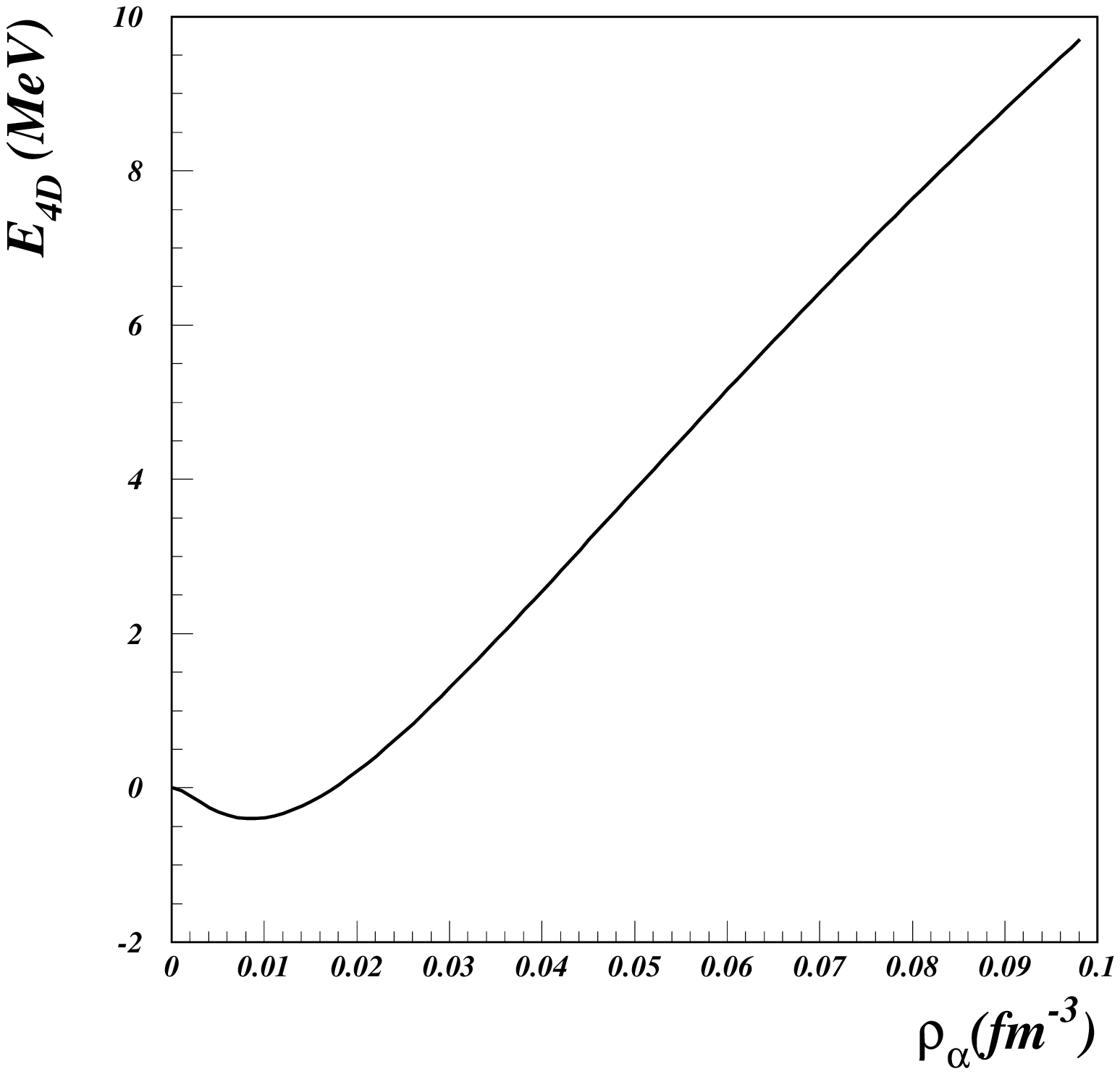}}
\caption{Density dependence of the ${\it E}_{\rm 4D}$ component.
\label{fig7}}
\end{figure}
and  the $D_i$'s are given by,
\beqa
D_i&=&-9\left [ \chi_{1,1}(\beta,\mu_i)+\chi_{3,1}(\beta,\mu_i)-2\chi_{2,1}(\beta,\mu_i)
-4(\chi_{6,5}(\beta,\mu_i)+\chi_{16,5}(\beta,\mu_i))\right.\nn\\
&&+8(\chi_{3,4}(\beta,\mu_i)+\chi_{11,4}(\beta,\mu_i)+\chi_{11,5}(\beta,\mu_i))\nn\\
&&-16(\chi_{2,3}(\beta,\mu_i)+\chi_{8,3}(\beta,\mu_i)
+\chi_{7,4}(\beta,\mu_i))\nn\\
&&+2^5(\chi_{5,3}(\beta,\mu_i)+\sqrt{2}(\chi_{22,7}(\beta,\mu_i) +\sqrt{2} \chi_{8,7}(\beta,\mu_i)))\nn\\
&&-2^6\sqrt{2}\chi_{15,7}(\beta,\mu_i)\nn\\
&&-2^7(\chi_{10,9}(\beta,\mu_i)+
\chi_{28,9}(\beta,\mu_i))\nn\\
&&+2^8(\chi_{6,7}(\beta,\mu_i)+\chi_{5,7}(\beta,\mu_i)+\chi_{19,9}(\beta,\mu_i)
+\chi_{19,7}(\beta,\mu_i)+\chi_{20,7}(\beta,\mu_i))\nn\\
&&-2^8(\chi_{8,11}(\beta,\mu_i)+\chi_{10,11}(\beta,\mu_i)
+\chi_{30,11}(\beta,\mu_i)+\chi_{32,11}(\beta,\mu_i))\nn\\
&&+2^9(\chi_{5,8}(\beta,\mu_i)+\chi_{21,8}(\beta,\mu_i)+\chi_{19,11}(\beta,\mu_i)
+\chi_{21,11}(\beta,\mu_i))\nn\\
&&-2^9(\chi_{3,5}(\beta,\mu_i)+\chi_{13,5}(\beta,\mu_i)
+\chi_{12,7}(\beta,\mu_i)+\chi_{13,7}(\beta,\mu_i))\nn\\
&&\left.+2^{10}(\chi_{8,5}(\beta,\mu_i)-\chi_{13,8}(\beta,\mu_i)) \right ] \ ,
\eeqa
where $(i=A,R)$.
\vspace{1cm}

{\subsubsection{Calculation of $E_{\rm 4O}$ open diagram} 
\label{E4O}}
Performing   the usual manipulations on the $E_{\rm 4O}$ term we obtain,\\
\beqa E_{\rm 4O}&=&\frac{1}{4}\rho^3\int \prod_1^5d\bd{R}_if^2(R_1)v^*(R_1)h(R_2)h(R_3)h(R_4)h(R_5)\nn\\
&&\delta(\bd{R}_1+\bd{R}_3+\bd{R}_4) \delta(\bd{R}_5-\bd{R}_1-\bd{R}_2) \ .
\eeqa
The latter equation is simply,
\beq E_{\rm 4O}=\frac{1}{4}\rho^3\int d\bd{R}_1f^2(R_1)v^*(R_1){[V_{h,h}^\delta(R_1)]}^2 \ ,\eeq
and after substitution of eq.(\ref{eq1}),
\beqa E_{\rm 4O}=\frac{\pi ^{9/2} \rho ^3}{\beta ^6} \left(\frac{c}{\beta}\left(-\frac{48}{19\sqrt{19}}+\frac{336}{169\sqrt{13}}+\frac{63-64\sqrt{3}}{294\sqrt{7}}+\right.\right.\nn\\ \left.\left.\frac{20\sqrt{6}-27}{150\sqrt{5}}-
\frac{3}{22\sqrt{11}}-\frac{177}{256\sqrt{2}}+\frac{3}{1024}+\frac{191\sqrt{3}}{576}\right)+P_RV_R-P_AV_A\right)\eeqa
where,
\beqa
P_i&=&\frac{1}{2} \chi_{1,1}(\beta,\mu_i)-\frac{255}{256}\chi_{2,1}(\beta,\mu_i)+\frac{63}{128}\chi_{3,1}(\beta,\mu_i)\nn\\&~&
+\frac{1}{256}\chi_{4,1}(\beta,\mu_i)+\frac{1}{4}\chi_{3,2}(\beta,\mu_i)-\frac{1}{2}\chi_{5,2}(\beta,\mu_i)\nn\\&~&+\frac{1}{4}\chi_{7,2}(\beta,\mu_i)+
\frac{4}{3\sqrt{3}} \chi_{4,3}(\beta,\mu_i)-\frac{1}{4}\chi_{5,3}(\beta,\mu_i)\nn\\&~&-\frac{8}{3\sqrt{3}}\chi_{7,3}(\beta,\mu_i)
+\frac{1}{2}\chi_{8,3}(\beta,\mu_i)+
\frac{4}{3 \sqrt{3}}\chi_{10,3}(\beta,\mu_i)\nn\\&~&-\frac{1}{4}\chi_{11,3}(\beta,\mu_i)-8\chi_{7,6}(\beta,\mu_i)+
16\chi_{13,6}(\beta,\mu_i)\nn\\&~&-8\chi_{19,6}(\beta,\mu_i)\nn\eeqa
\begin{figure}[h]
\centerline{\includegraphics[width=10cm]{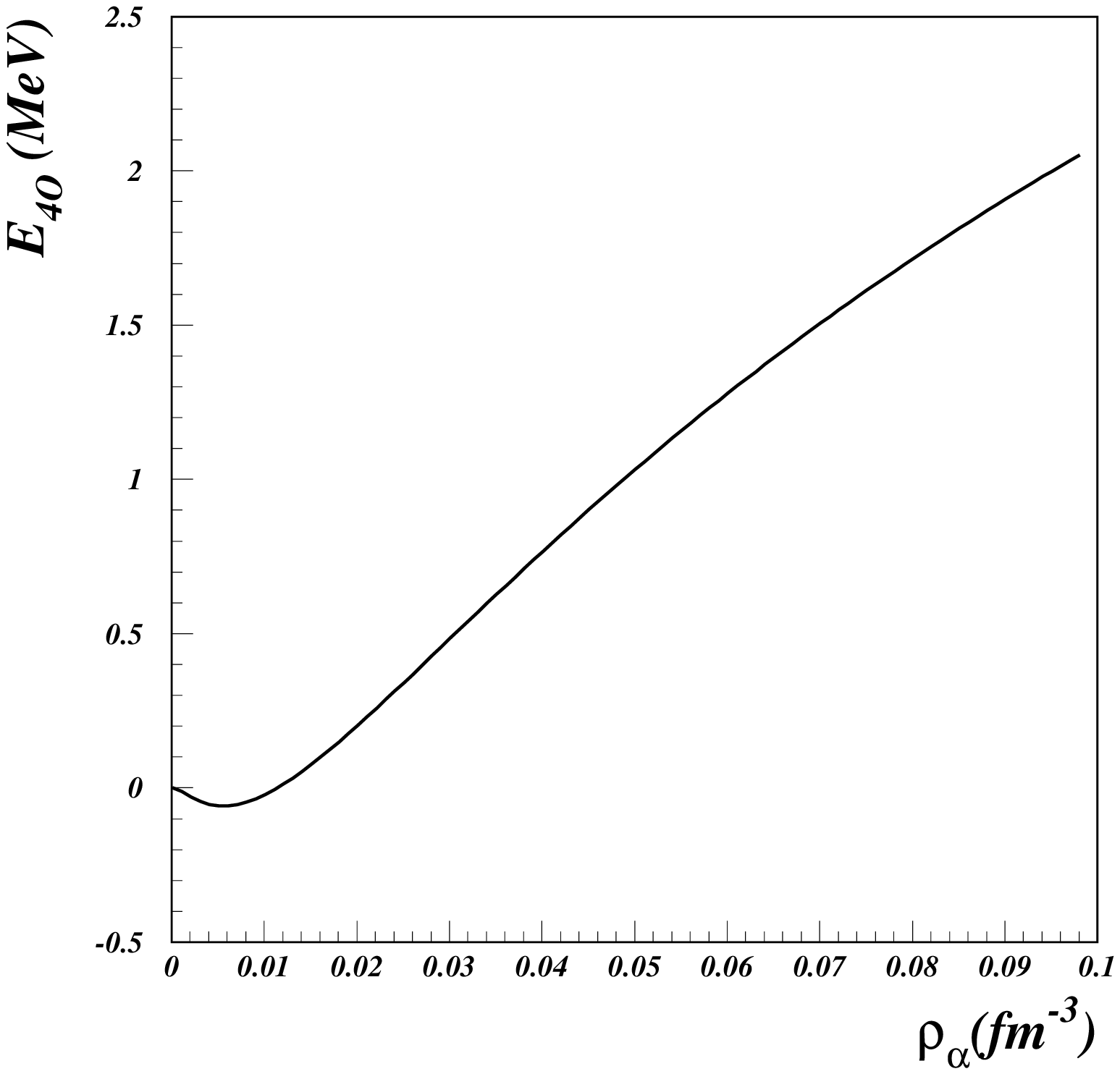}}
\caption{Density dependence of the ${\it E}_{\rm 4O}$
 component.
\label{fig8}}
\end{figure}
\vspace{1cm}

The density dependence of the $E_{4O}$ component is displayed in Fig.
\ref{fig8}.

{\subsubsection{Calculation of $E_{4C}$ connected diagram} \label{e4c}}
In this subsection we  estimate,
\beq E_{\rm 4C}=\frac{1}{4}\frac{\rho^3}{\Omega}\int \prod_1^4 d\bd{r}_i v^*(r_{12})f^2(r_{12})h(r_{23})h(r_{34})h(r_{41})h(r_{24})h(r_{13}) \ .\eeq
Using the same  technique as above we introduce $\delta$ kernels,
\beqa E_{\rm 4C}=\frac{1}{4}\rho^3 \int \prod_1^6d\bd{R}_iv^*(R_1)f^2(R_1)h(R_2)h(R_3)h(R_4)h(R_5)h(R_6)\nn\\\delta(\bd{R}_6+\bd{R}_3+\bd{R}_4) \delta(\bd{R}_5-\bd{R}_2-\bd{R}_3)\delta(\bd{R}_6-\bd{R}_1-\bd{R}_2) \ .
\label{eq81}
\eeqa
The identities
\beq
(2 \pi)^3  \delta\left(\bd{R_6}+\bd{R_3}+\bd{R_4}\right)=\int d\bd{q_1} \exp(\bd{q_1}.(\bd{R_6}+\bd{R_3}+\bd{R_4})) \ ,
\eeq
\beq
(2 \pi)^3 \delta\left(\bd{R_5}-\bd{R_2}-\bd{R_3}\right)=\int d\bd{q_2} \exp(\bd{q_2}.(\bd{R_5}-\bd{R_2}-\bd{R_3}))  ,
\eeq
\beq
(2 \pi)^3 \delta\left(\bd{R_6}-\bd{R_1}-\bd{R_2}\right)=\int d\bd{q_3} \exp(\bd{q_3}.(\bd{R_6}-\bd{R_1}-\bd{R_2})) \ ,
\eeq
 are introduced  in Eq.(\ref{eq81}). The latter equation is then expressed in terms of the Fourier transform of $h$, labelled $\widetilde h$ and  that of $v^* f^2$,
 labelled $G$. More precisely,
\beq
E_{\rm 4C}=\frac{1}{2048 \pi^9} \rho^3 \int \prod_{i=1}^3 d\bd{q_ i} \ \widetilde h(\bd{q_1}-\bd{q_2}) \widetilde h(\bd{q_1}+\bd{q_3}) \widetilde h(\bd{q_2}+\bd{q_3})
 \widetilde h(\bd{q_1}) \widetilde h(\bd{q_2})  G(\bd{q_3}) \ .
\label{E4cM}
\eeq
For the sake of simplicity use is made of the scaling  $\bd{q_j} \mapsto 2 \sqrt{2} \beta \bd{q_j}$ in
Eq.(\ref{E4cM}) in order to eliminate the $\beta$ dependence in $\widetilde h$.
Introducing,
\beq
 h^*(\bd{q})= -2 e^{-2 q^2} +\frac{1}{2 \sqrt{2}}  e^{-q^2} \ ,
\eeq 
 such that  $\widetilde h(2 \sqrt{2} \beta \bd{q}) \equiv  (\sqrt{\pi}/b)^3 h^*( \bd{q})$,
the equation (\ref{E4cM}) becomes,
\beqa
E_{\rm 4C}=\frac{4 \sqrt{2} \ }{\beta^6 \sqrt{\pi}^3 } \rho^3 \int \prod_{i=1}^3 d\bd{q_ i} \   h^*(\bd{q_1}-\bd{q_2}) h^*(\bd{q_1}+\bd{q_3})  h^*(\bd{q_2}+\bd{q_3})\nn\\
h^*(\bd{q_1})  h^*(\bd{q_2})  G( 2 \sqrt{2} \beta \bd{ q_3}) \ .
\eeqa
The auxiliary integral 
\beq
J(\bd{q_3})=\int \prod_{i=1}^2 d\bd{q_ i} \   h^*(\bd{q_1} - \bd{q_2}) h^*(\bd{q_1}+ \bd{q_3})  h^*(\bd{q_2}+\bd{q_3})
 h^*(\bd{q_1}) h^*(\bd{q_2})   \ ,
\eeq
entering the definition of $E_{4c}$, has been performed by using the Cartesian coordinates. We obtain after calculations 
$J(\bd{q_3})=\pi^3 \ Q(q_3)$ with, 
\begin{eqnarray}
Q(q) &=&\frac{-9 \ \sqrt{2} + 2 \ \sqrt{3}}{72} \ \exp(-2 \ q^2) + \frac{8}{13 \ \sqrt{13}} \ \exp\left(-\frac{22}{13} \ q^2 \right) \nonumber\\
 & -&  \frac{  4}{19 \ \sqrt{19}} \  \exp\left(-\frac{32}{19} \ q^2 \right) + \left( \frac{ 1}{56 \ \sqrt{7}}-\frac{1}{20 \ \sqrt{5}} \right) 
   \exp\left(-\frac{3}{2} q^2 \right) \nonumber\\
 & -& \frac{2}{21 \ \sqrt{21}}  \  \exp\left(-\frac{10}{7} q^2 \right) + 
  \frac{1}{30 \ \sqrt{30}}  \  \exp\left(-\frac{7}{5} q^2 \right) \nonumber\\
 & +&     \left( \frac{ 1}{30 \ \sqrt{30}}-\frac{2}{21 \ \sqrt{21}} \right)     \exp\left(-\frac{4}{3} \ q^2 \right)                           
   + \frac{1}{64 \ \sqrt{2}}      \exp\left(-\frac{19}{16} \ q^2 \right)      \nonumber\\       
 & - &  \frac{1}{88 \ \sqrt{11}}    \exp\left(-\frac{13}{11} \ q^2 \right) + \frac{ 9 - 16 \ \sqrt{3}}{36864}  \ \exp(-q^2) \ .
\end{eqnarray}

\begin{figure}[h]
\centerline{\includegraphics[width=10cm]{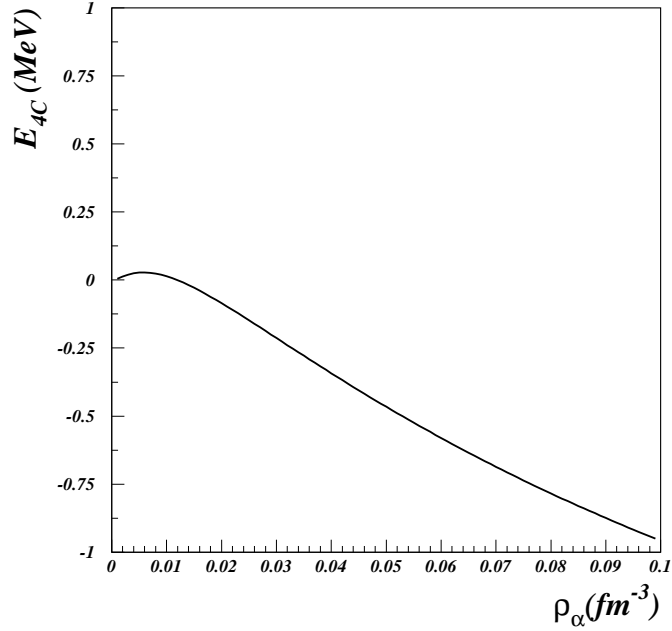}}
\caption{Density dependence of the ${\it E}_{\rm 4C}$ component.}
\label{fig9}
\end{figure}

Therefore,
\beq
E_{\rm 4C}=\frac{4 \sqrt{2} \sqrt{\pi}^3 \ }{\beta^6 \ } \rho^3 \int \ d\bd{q_ 3} \   Q(q_3)  G( 2 \sqrt{2} \beta \bd{ q_3})
\eeq
is expressed in terms of  the 3D Fourier transform ($G$) of $v^*(r) f^2(r)$, 
 which is  given by, 

\begin{figure}[h]
\centerline{\includegraphics[width=10cm]{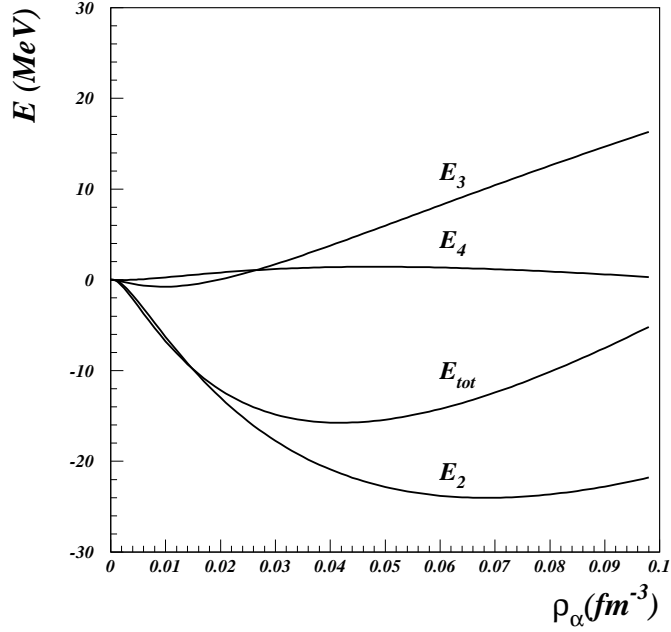}}
\caption{${\it E}_2,{\it E}_3,{\it E}_4,{\it E_{tot}}$ density dependence.}
\label{fig10}
\end{figure}

\begin{eqnarray}
G(q) &=& 
\pi^{3/2} \ \left\{ c \ \left[ \frac{3}{\sqrt{2 \ } \ \beta} \  \exp \left( -\frac{q^2}{8 \ \beta^2} \right) - 
   \frac{q^2}{\beta^3} \  \exp \left(-\frac{q^2}{4 \ \beta^2} \right) \right] \right. \nonumber\\ & & \left.
  +  V_R \left[ \frac{1}{4 \ \mu_R^3}  \exp \left( -\frac{q^2}{4 \mu_R^2} \right) -
   \frac{2}{(\beta^2 + \mu_R^2 \ )^{3/2}}  \exp \left(-\frac{q^2}{4 \ (\beta^2 + \mu_R^2) }\right) \right. \right. \nonumber\\ 
 & & \left. \left.  + \frac{1}{ (2 \ \beta^2 + \mu_R^2 \ )^{3/2}}  \exp \left(-\frac{q^2}{4 \ (2 \ \beta^2 + \mu_R^2) }\right) \right]  
  -  V_A \left[ \frac{1}{4 \ \mu_A^3}  \exp \left( -\frac{q^2}{4 \mu_A^2} \right) \right.\right. \nonumber\\
 & &   - \frac{2}{ (\beta^2 + \mu_A^2 \ )^{3/2}}  \exp \left(-\frac{q^2}{4 \ (\beta^2 + \mu_A^2) }\right) \nn\\ & &\left.\left.
  + \frac{1}{ (2 \ \beta^2 + \mu_A^2 \ )^{3/2}}  \exp \left(-\frac{q^2}{4 \ (2 \ \beta^2 + \mu_A^2) }\right) \right] \right \} \ ,
\nonumber
\end{eqnarray}
we are left with the final expression of $E_{4C}$:
\begin{eqnarray}
E_{\rm 4C}&=&\frac{4 \sqrt{2} \ \pi^{9/2} }{\beta^6  } \rho^3 \left[\frac{c}{\beta}
\left(-\frac{3142463}{276480000} - \frac{200 \ \sqrt{2}}{867 \ \sqrt{51} } + \frac{657 \ 
\sqrt{2}}{1225 \ \sqrt{35} } \right. \right. \nonumber\\
&+& \left. \left. \frac{239063}{2074464 \ \sqrt{2}} - \frac{19}{64 \ \sqrt{3}} + \frac{243}{2450 \ \sqrt{35}} \right) 
   +C_R V_R - C_A V_A \right] \ .
\end{eqnarray}
The latter equation is expressed in terms of the $C_i$'s, $i=A,R$ which are listed in the Appendix.

\begin{figure}[t]
\center{\epsfig{figure=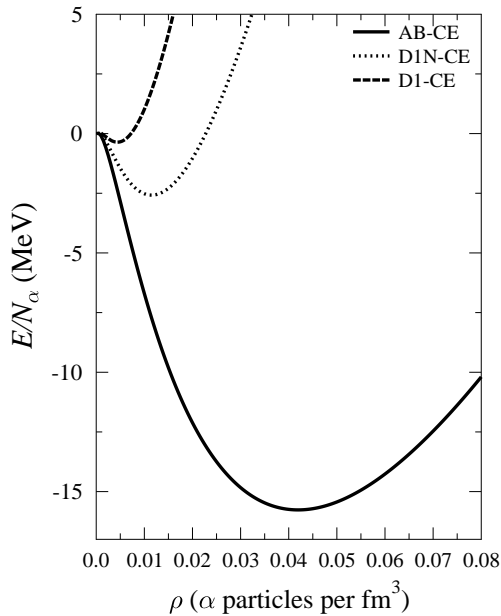,width=0.5\textwidth}}
\caption{Total energy for the case of the Ali-Bodmer potential compared to the
Gogny D1 and D1N}
\label{fig11}
\end{figure}

 The contribution to the  energy supplied by each component of the cluster
 expansion is  summarized in Fig.\ref{fig10} for the
 AB interaction.
 Clearly the major contributions come from $E_2$ and $E_3$ whereas 
 the four-body contributions are very weak. In contrast to two-body 
 contributions the three-body and four-body 
contribution are repulsive and the equilibrium arises from a delicate 
balance between $E_2$ and $E_3$ components.

\begin{figure}[h]
\centering
\mbox{\subfigure{\epsfig{figure=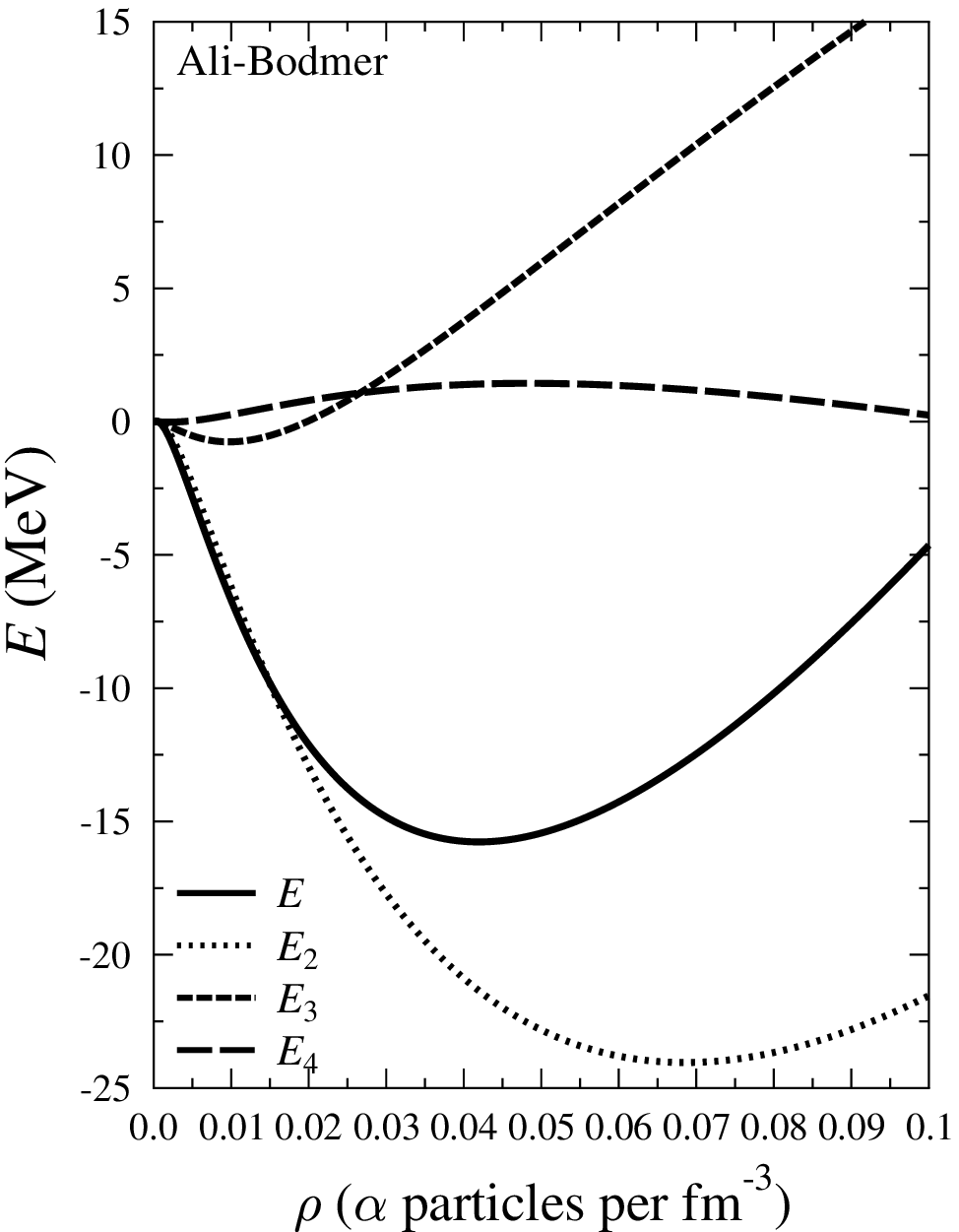,width=0.32\textwidth}}
      \subfigure{\epsfig{figure=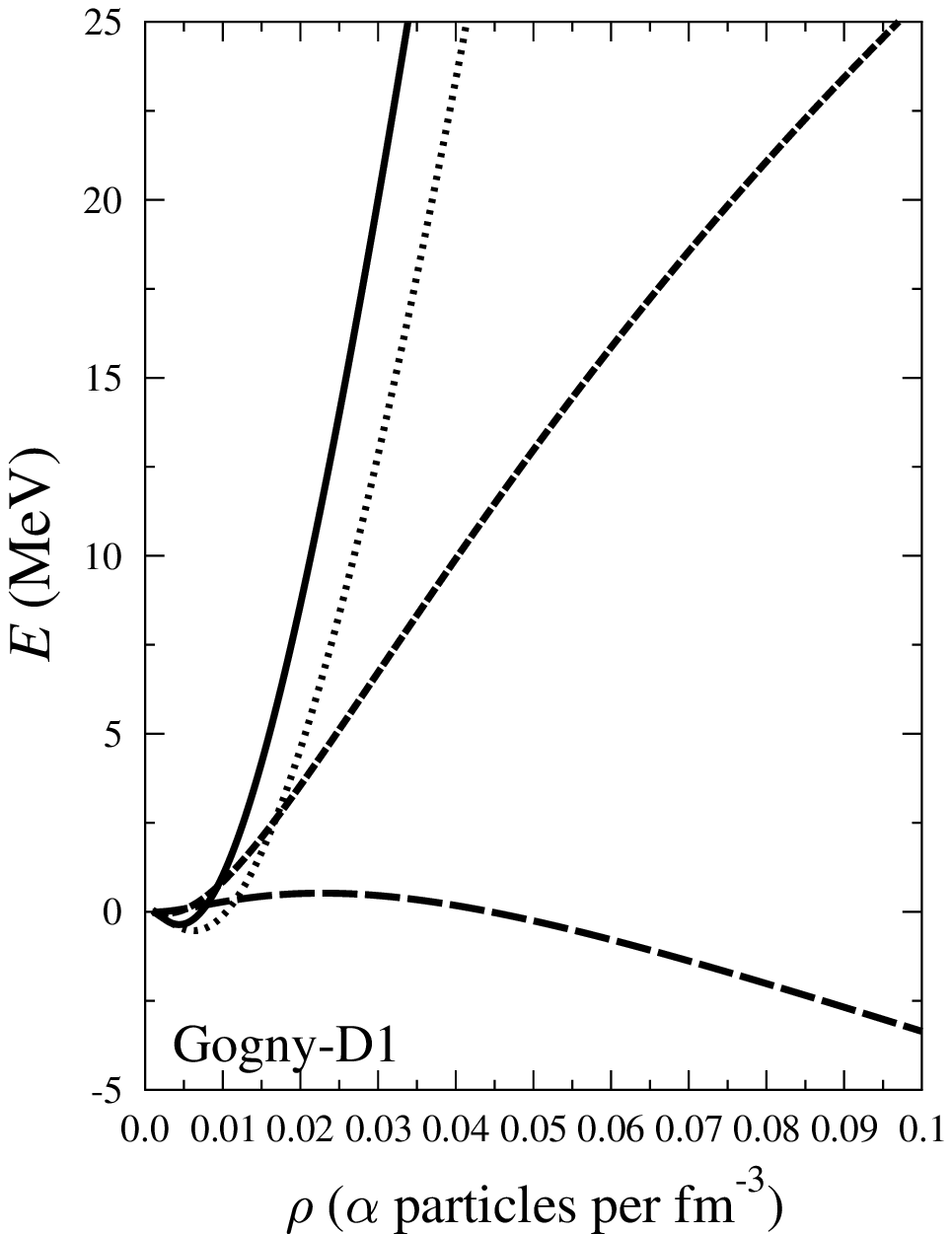,width=0.32\textwidth}}
      \subfigure{\epsfig{figure=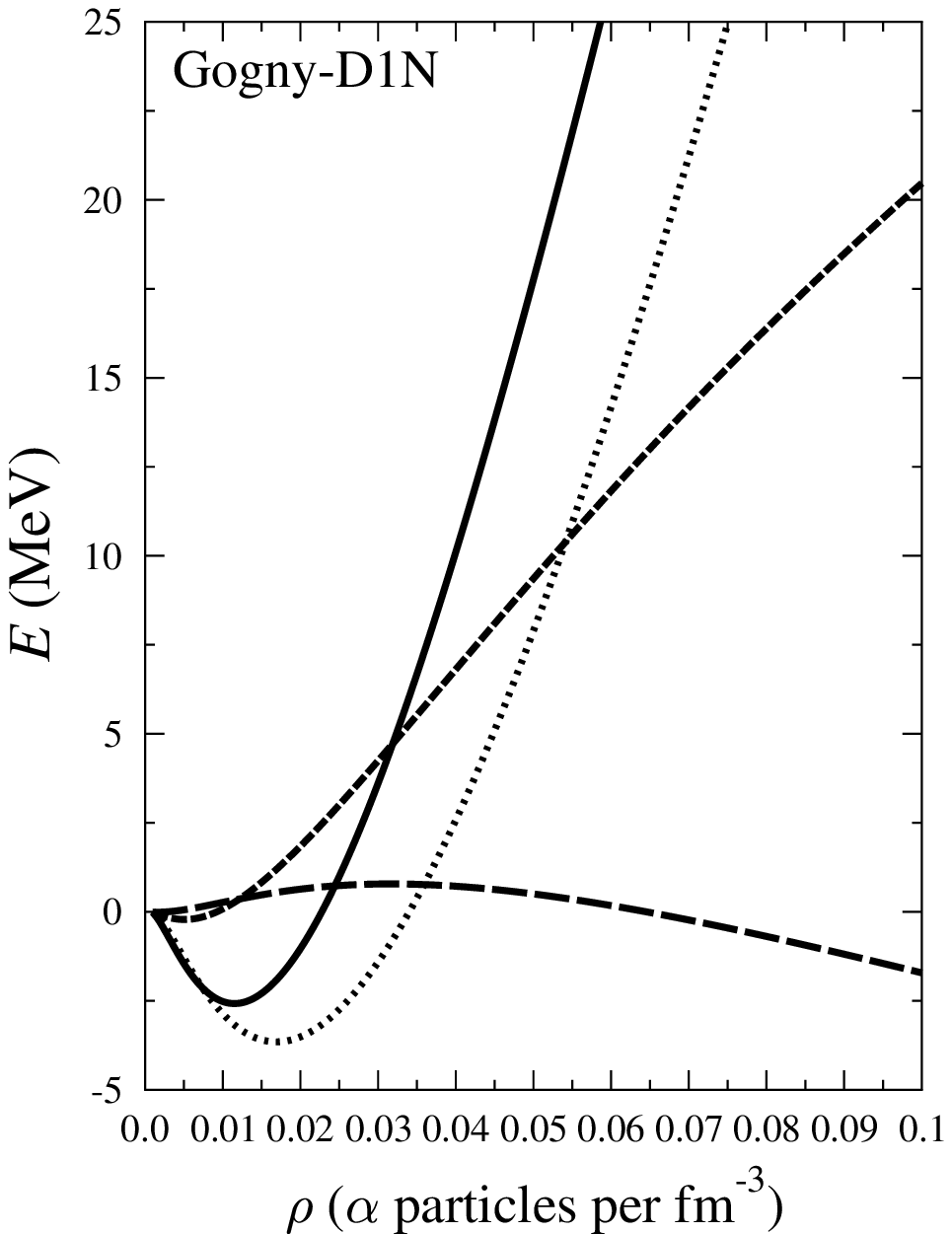,width=0.32\textwidth}}
      } 
       
\caption{${\it E}_2,{\it E}_3,{\it E}_4,{\it E}$ dependence of the 
$\alpha$-matter density $\rho$.(Ali-Bodmer and Gogny) 
}
\label{fig12}
\end{figure}

We  display in Fig. \ref{fig11} a comparison of the EOS determined using 
the Ali-Bodmer potential with 
EOS obtained using D1 and D1N parametrizations of the Gogny potential. More
details are given in Fig. \ref{fig12}. While
the equilibrium point predicted by the AB interaction is deep and lies at the
normal nuclear matter density, Gogny parametrizations predict a shallower
minimum at densities close to the Mott density.


\begin{figure}[h]
\centering
\mbox{\subfigure{\epsfig{figure=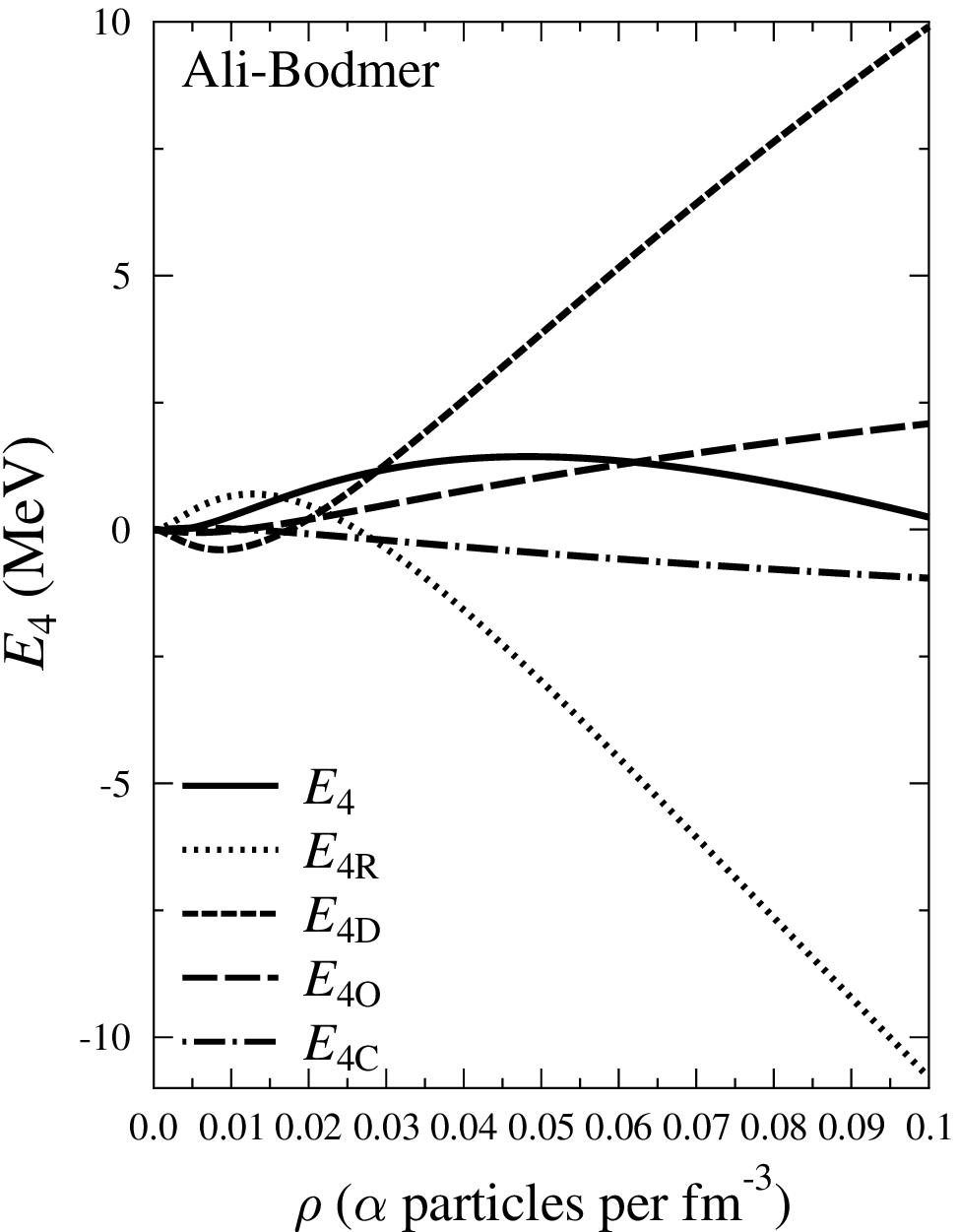,width=0.32\textwidth}}
      \subfigure{\epsfig{figure=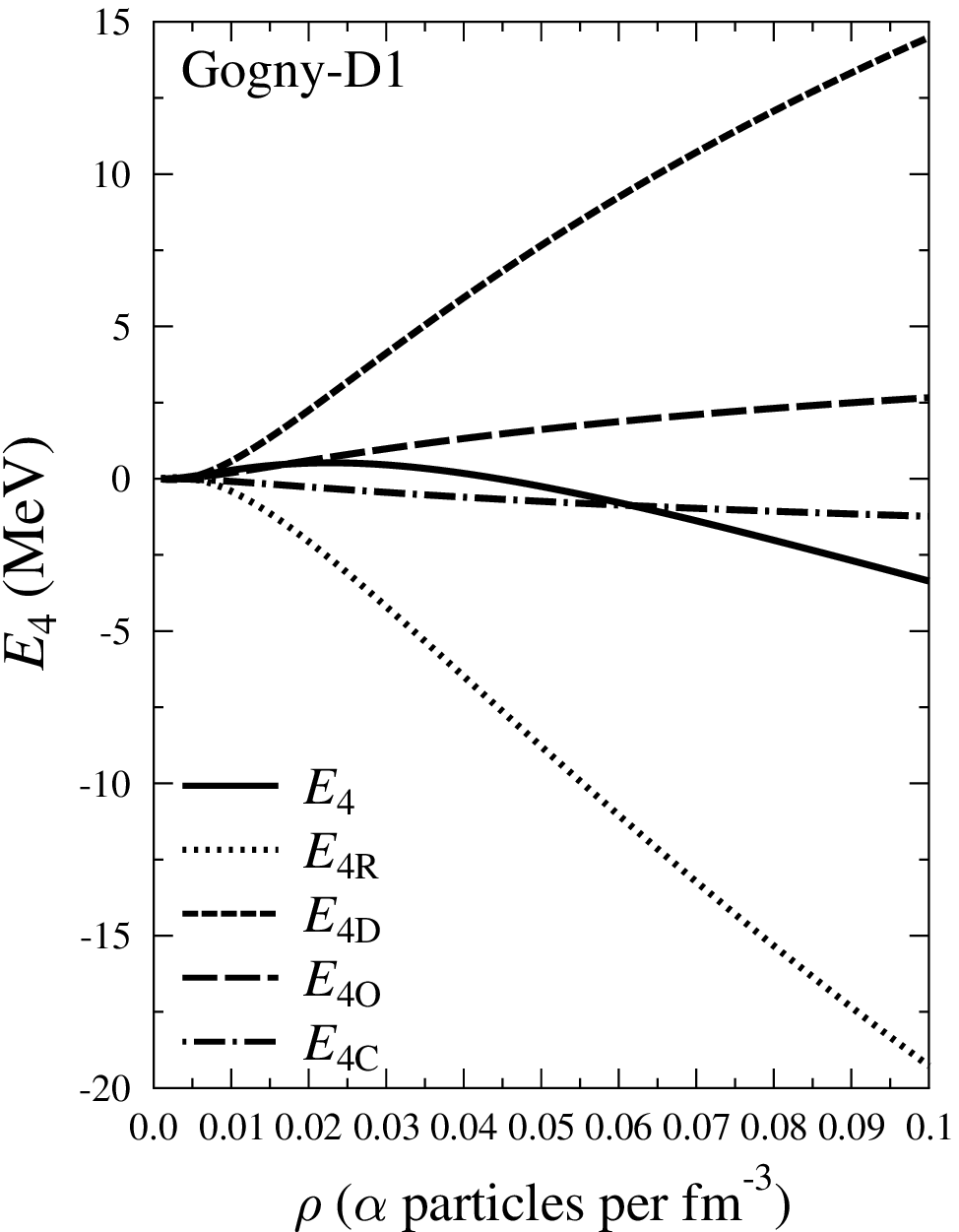,width=0.32\textwidth}}
      \subfigure{\epsfig{figure=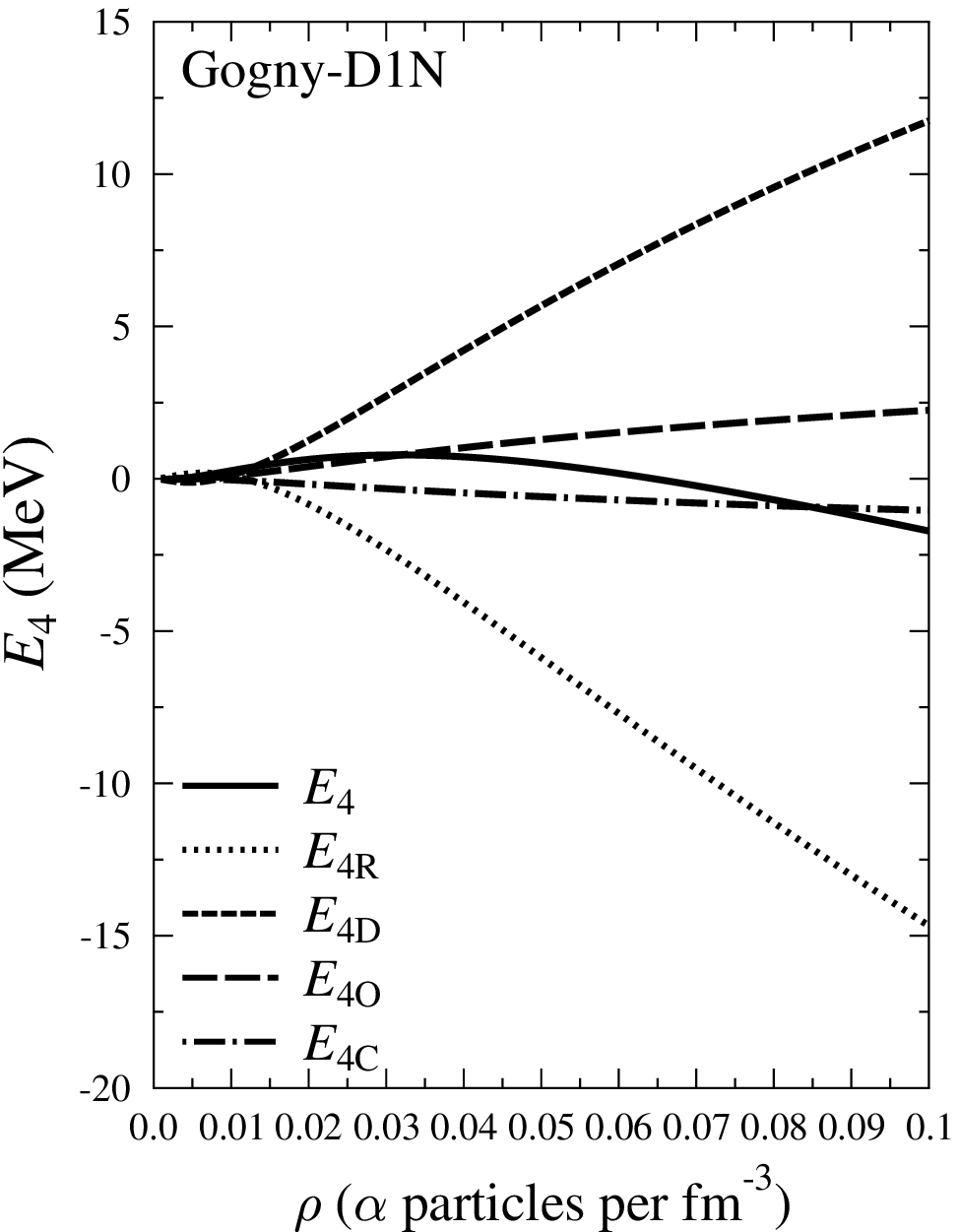,width=0.32\textwidth}}
} 
       
\caption{${\it E}_{\rm 4R},{\it E}_{\rm 4D},{\it E}_{\rm 4O},{\it E}_{\rm 4C},{\it E}_4$ dependence of the 
$\alpha$-matter density $\rho$.(Ali-Bodmer and Gogny) 
}
\label{fig13}
\end{figure}

The four-body contribution to the total energy is small compared to other
components. This is explained by an almost complete cancellation between
$E_{4R}$ and $E_{4D}$ while the most connected diagrams $E_{4O}$ and $E_{4C}$ are
intrinsically small due to increased number of bonds (see Fig. \ref{fig13}).

\section{Concluding remarks}

By assuming a simple but realistic functional form for the TBCFN's and employing 
Gaussian-like potentials, we obtained analytical expressions for the first six terms 
(one corresponding to the two-body, one to the three-body and four for the four-body diagrams) occurring in the cluster expansion of an extended uniform system of structureless, indistinguishable $\alpha$ particles in a uniform background of neutralizing charge.
The results are pointing to a saturation of the alpha matter at densities around the
nuclear matter density (for Ali-Bodmer potential) and for the D1 interaction at almost one 
tenth of this density.     
We also demonstrated the strength of the folding method, previously applied
in the calculation of the heavy-ion potential \cite{carst92} to the calculation 
of cluster integrals of a many-boson system. 
\newpage
\section{Appendix}

In this section we report the coefficients $C_i$ entering the final expression of $E_{4C}$. They are 
given below in terms of the coefficients,
\beq
\chi_{mn}=\frac{1}{(m \beta^2+n \mu_i^2)^{3/2}} \ .
\eeq
We have,
\begin{eqnarray}
 C_i&=&\left(\frac{1}{24 \ \sqrt{6}} -\frac{1}{16}\right) \ \chi_{1, 1}(\beta,\mu_i)   + \left( \frac{513}{4096}-
\frac{1}{768 \ \sqrt{3}}-\frac{1}{ 12 \ \sqrt{6}} \right) \ \chi_{2, 1}(\beta,\mu_i)) \nonumber\\
  &+& \left( -\frac{129}{2048}+ \frac{1}{384 \ \sqrt{3}}+\frac{1}{24 \ \sqrt{6}} \right) \ \chi_{3, 1}(\beta,\mu_i)\nn\\ 
  &+ & \left(\frac{1}{40 \ \sqrt{5}}  - \frac{1}{7 \ \sqrt{14}} \right)  \chi_{3, 2}(\beta,\mu_i) \nonumber\\ 
&+&  \left( \frac{1}{4096} -\frac{1}{768 \ \sqrt{3}} \right) \  \chi_{4, 1}(\beta,\mu_i)  
  +   \left( -\frac{1}{ 5 \ \sqrt{10}}+\frac{1}{14 \ \sqrt{14}} \right) \ \chi_{4, 3}(\beta,\mu_i) \nonumber\\
  & +& \left(  \frac{\sqrt{2}}{7 \sqrt{7}}-\frac{1}{20 \ \sqrt{5}} \right) \ \chi_{5, 2}(\beta,\mu_i)  
 + \left( \frac{1}{40 \ \sqrt{5}} -  \frac{1}{ 7 \ \sqrt{14}} \right)  \ \chi_{7, 2}(\beta,\mu_i) \nonumber\\
  & + & \left( \frac{\sqrt{2}}{ 5 \ \sqrt{5}} - \frac{1}{7 \ \sqrt{14}} \right)  \  \chi_{7, 3}(\beta,\mu_i) 
  -  \frac{1}{3 \ \sqrt{6}} \  \chi_{7, 5}(\beta,\mu_i) \nonumber\\
 & + & \left( -\frac{1}{5 \ \sqrt{10}}+\frac{1}{14 \ \sqrt{14}} \right) \ \chi_{10, 3}(\beta,\mu_i) 
  +\frac{1}{6 \ \sqrt{6}} \  \chi_{10, 7}(\beta,\mu_i) \nonumber\\
 & + &\frac{ \sqrt{2}}{3 \sqrt{3}}  \ \chi_{12, 5}(\beta,\mu_i)
  + 2 \ \sqrt{2} \ \chi_{13, 11}(\beta,\mu_i) - \frac{1}{3 \ \sqrt{6}}  \chi_{17, 5}(\beta,\mu_i) \nonumber\\
 &-& \frac{1}{3 \ \sqrt{6}} \  \chi_{17, 7}(\beta,\mu_i) - \sqrt{2} \ \chi_{19, 16}(\beta,\mu_i)  
 - \frac{1}{8} \ \chi_{22, 13}(\beta,\mu_i) \nonumber\\
 &+& \frac{1}{6 \ \sqrt{6}} \  \chi_{24, 7}(\beta,\mu_i) 
  - 4 \ \sqrt{2} \ \chi_{24, 11}(\beta,\mu_i) + \frac{1}{\sqrt{2}} \ \chi_{32, 19}(\beta,\mu_i) \nonumber\\
& + &  2 \ \sqrt{2} \ \chi_{35, 11}(\beta,\mu_i) + \frac{1}{4} \ \chi_{35, 13}(\beta,\mu_i) 
 + 2 \ \sqrt{2} \ \chi_{35, 16}(\beta,\mu_i)\nn\\& -& \frac{1}{8} \ \chi_{48, 13}(\beta,\mu_i) 
-   \sqrt{2} \ \chi_{51, 16}(\beta,\mu_i) \nn\\&- &\sqrt{2} \ \chi_{51, 19}(\beta,\mu_i) + \frac{1}{\sqrt{2}} \ 
\chi_{70, 19}(\beta,\mu_i). 
\end{eqnarray}
where $(i=A,R)$.

\newpage

\begin{center}

\end{center}

{\centering\subsubsection*{Acknowledgements}}
We are indebted to Roland Lombard for reading the manuscript and for suggestions. 
This work was partly supported by CNCSIS Romania,
under program PN-II-PCE-2007-1, contracts No.49 and No.258 and by CNMP contract PNCDI2 D7 7.4 No.71112.

\end{document}